%% file: paper-introspection.tex
\documentclass[sigconf,authorversion=true]{acmart}

\usepackage{booktabs} % For formal tables

\usepackage[utf8]{inputenc}
\usepackage{multirow}
\usepackage{microtype}
\usepackage{todonotes}

\usepackage{subcaption}

\usepackage[toc,acronym,nowarn]{glossaries}
\input{gloss-acr}
\makeglossaries
\glsdisablehyper % disable hyperlinks to nowhere...
\usepackage{xspace}

\newcommand{\Title}{Extracting Secrets from Encrypted Virtual Machines}

\usepackage{graphicx}
\DeclareGraphicsExtensions{.pdf,.jpeg,.png}

\usepackage{hyperref}
\hypersetup{
  colorlinks=true, linkcolor=black, citecolor=black, urlcolor=black,
  pdftitle={\Title}
  pdfauthor={Mathias Morbitzer, Manuel Huber, Julian Horsch}
}

\usepackage[all]{hypcap}
\usepackage{fixltx2e}
\usepackage{enumitem} % fixes overfull hboxes in description envs
\usepackage{nicefrac}
\usepackage{amssymb}
\usepackage{pbox}

\begin{document}

\copyrightyear{2019}
\acmYear{2019}
\setcopyright{acmlicensed}
\acmConference[CODASPY '19]{Ninth ACM Conference on Data and Application Security and Privacy}{March 25--27, 2019}{Richardson, TX, USA}
\acmBooktitle{Ninth ACM Conference on Data and Application Security and Privacy (CODASPY '19), March 25--27, 2019, Richardson, TX, USA}
\acmPrice{15.00}
\acmDOI{10.1145/3292006.3300022}
\acmISBN{978-1-4503-6099-9/19/03}

\title{\Title}

\author{Mathias Morbitzer}
%\orcid{}
\authornote{Both authors contributed equally to the work.}
\affiliation{
  \institution{Fraunhofer AISEC}
  \city{Garching near Munich}
  \state{Germany}
}
\email{morbitzer@aisec.fraunhofer.de}

\author{Manuel Huber}
\orcid{0000-0003-0829-6902}
\authornotemark[1]
\affiliation{
  \institution{Fraunhofer AISEC}
  \city{Garching near Munich}
  \state{Germany}
}
\email{manuel.huber@aisec.fraunhofer.de}

\author{Julian Horsch}
\orcid{0000-0001-9018-7048}
\affiliation{
  \institution{Fraunhofer AISEC}
  \city{Garching near Munich}
  \state{Germany}
}
\email{julian.horsch@aisec.fraunhofer.de}

\begin{abstract}
\input{abstract}
\end{abstract}




\maketitle

\newpage
\input{content}

\bibliographystyle{ACM-Reference-Format}
\bibliography{biblio}

\end{document}

%% file: gloss-acr.tex
\newacronym{OS}{OS}{Operating System}
\newacronym{SE}{SE}{Secure Element}
\newacronym{NFC}{NFC}{Near Field Communication}
\newacronym{BLE}{BLE}{Bluetooth Low Energy}
\newacronym{RIL}{RIL}{Radio Interface Layer}
\newacronym{SELinux}{SELinux}{Security-Enhanced Linux}
\newacronym{LSM}{LSM}{Linux Security Modules}
\newacronym{TCB}{TCB}{Trusted Computing Base}
\newacronym{MAC}{MAC}{Mandatory Access Control}
\newacronym{IPC}{IPC}{Inter-Process Communication}
\newacronym{ICC}{ICC}{Inter-Container Communication}
\newacronym{cgroups}{cgroups}{control groups}
\newacronym{CA}{CA}{Certificate Authority}
\newacronym{PKI}{PKI}{Public Key Infrastructure}
\newacronym{MITM}{MITM}{Man-In-The-Middle}
\newacronym{BYOD}{BYOD}{Bring-Your-Own-Device}
\newacronym{CM}{CM}{Container Management}
\newacronym{SM}{SM}{Security Management}
\newacronym{HAL}{HAL}{Hardware Abstraction Layer}
\newacronym{TLS}{TLS}{Transport Layer Security}
\newacronym{C2C}{C2C}{Container To Container}
\newacronym{Protobuf}{Protobuf}{Protocol Buffers}
\newacronym{SDO}{SDO}{Sensitive Data Object}
\newacronym{VMA}{VMA}{Virtual Memory Area}
\newacronym{PGD}{PGD}{Page Global Directory}
\newacronym{PUD}{PUD}{Page Upper Directory}
\newacronym{PMD}{PMD}{Page Middle Directory}
\newacronym{PTE}{PTE}{Page Table Entry}
\newacronym{COW}{COW}{Copy-On-Write}
\newacronym{IV}{IV}{Initialization Vector}
\newacronym{ESSIV}{ESSIV}{Encrypted Salt-Sector Initialization Vector}
\newacronym{KSM}{KSM}{Kernel Samepage Merging}
\newacronym{JIT}{JIT}{Just-In-Time}
\newacronym{DMA}{DMA}{Direct Memory Access}
\newacronym{FDE}{FDE}{Full Disk Encryption}
\newacronym{AS}{AS}{Address Space}
\newacronym{GCM}{GCM}{Google Cloud Messaging}
\newacronym{TPM}{TPM}{Trusted Platform Module}
\newacronym{JTAG}{JTAG}{Joint Test Action Group}
\newacronym{LUKS}{LUKS}{Linux Unified Key Setup}
\newacronym{VPN}{VPN}{Virtual Private Network}
\newacronym{PBKDF2}{PBKDF2}{Password-Based Key Derivation Function 2}
\newacronym{KVM}{KVM}{Kernel-based Virtual Machine}
\newacronym{VM}{VM}{Virtual Machine}
\newacronym{HV}{HV}{Hypervisor}
\newacronym{SEV}{SEV}{Secure Encrypted Virtualization}
\newacronym{SME}{SME}{Secure Memory Encryption}
\newacronym{SP}{SP}{Secure Processor}
\newacronym[firstplural=Guest Virtual Adresses (GVAs)]{GVA}{GVA}{Guest Virtual Address}
\newacronym[firstplural=Guest Physical Addresses (GPAs)]{GPA}{GPA}{Guest Physical Address}
\newacronym[firstplural=Host Physical Addresses (HPAs)]{HPA}{HPA}{Host Physical Address}
\newacronym{GPT}{GPT}{Guest Page Table}
\newacronym{HPT}{HPT}{Host Page Table}
\newacronym{TLB}{TLB}{Translation Lookaside Buffer}
\newacronym{PoC}{PoC}{Proof of Concept}
\newacronym{ORAM}{ORAM}{Oblivious RAM}
\newacronym{SEV-ES}{SEV-ES}{SEV Encrypted State}
\newacronym{VMCB}{VMCB}{Virtual Machine Control Block}
\newacronym{SLAT}{SLAT}{Second Level Address Translation}
\newacronym{SSH}{SSH}{Secure Shell}
\newacronym{RSA}{RSA}{Rivest–Shamir–Adleman}
\newacronym{ECDHE}{ECDHE}{Elliptic-Curve Diffie-Hellman Ephemeral}
\newacronym{AES}{AES}{Advanced Encryption Standard}
\newacronym{OOM}{OOM}{Out Of Memory}
\newacronym{MKTME}{MKTME}{Multi-Key Total Memory Encryption}
\newacronym{VMI}{VMI}{Virtual Machine Introspection}
\newacronym{MAD}{MAD}{Median Absolute Deviation}
\newacronym{HSM}{HSM}{Hardware Security Module}

%% file: abstract.tex
AMD SEV is a hardware extension for main memory encryption on multi-tenant systems.
SEV uses an on-chip coprocessor, the AMD Secure Processor, to transparently encrypt virtual machine memory with individual, ephemeral keys never leaving the coprocessor.
The goal is to protect the confidentiality of the tenants' memory from a malicious or compromised hypervisor and from memory attacks, for instance via cold boot or DMA.
The SEVered attack has shown that it is nevertheless possible for a hypervisor to extract memory in plaintext from SEV-encrypted virtual machines without access to their encryption keys.
However, the encryption impedes traditional virtual machine introspection techniques from locating secrets in memory prior to extraction.
This can require the extraction of large amounts of memory to retrieve specific secrets and thus result in a time-consuming, obvious attack.
We present an approach that allows a malicious hypervisor quick identification and theft of secrets, such as TLS, SSH or FDE keys, from encrypted virtual machines on current SEV hardware.
We first observe activities of a virtual machine from within the hypervisor in order to infer the memory regions most likely to contain the secrets.
Then, we systematically extract those memory regions
and analyze their contents on-the-fly.
This allows for the efficient retrieval of targeted secrets, strongly increasing the chances of a fast, robust and stealthy theft.

%% file: content.tex
\section{Introduction}
\label{sec:introduction}

 On common multi-tenant systems, the confidentiality of sensitive \gls{VM} data depends on both the \glsdisp{HV}{Hypervisor's (HV)} integrity and on the operator's trustworthiness.
 Unfortunately, these strong requirements are prone to getting infringed by different attack vectors.
 Examples are attacks by other tenants exploiting software-level vulnerabilities to escape their sandboxed \glspl{VM} \cite{Microsoft2017Vuln, VMWare2017Vuln, Xen2017Vuln}, attackers with physical access conducting a memory attack, e.g., via \gls{DMA} \cite{boileau2006hit,pcie,becher2005firewire} or cold boot \cite{Halderman08lestwe}, or simply a malicious operator using the \gls{HV} to read the \gls{VM}'s memory.
 In order to protect the \gls{VM}'s memory in such scenarios, AMD introduced \gls{SEV} \cite{AMD2017API,sev-es} on recent server systems.
 \gls{SEV} is a hardware extension for main memory encryption on a per-\gls{VM} granularity.
 With \gls{SEV} enabled, AMD's \gls{SP} transparently encrypts the main memory of each \gls{VM} with individual \gls{SP}-bound keys.
 The goal is to protect \glspl{VM} from memory attacks and from a malicious or compromised \gls{HV}.
 To attest tenants that their \glspl{VM}' memory is indeed encrypted, \gls{SEV} includes a cryptographic protocol to verify \gls{VM} encryption on an \gls{SEV}-enabled platform.

 SEVered \cite{Morbitzer:2018:SSA:3193111.3193112} is a recent attack on AMD \gls{SEV}, which showed that it is nevertheless possible for a \gls{HV} to extract plaintext contents from \gls{SEV}-encrypted \glspl{VM}.
 SEVered exploits \gls{SEV}'s missing integrity protection for \gls{VM} memory pages, previously discovered by \cite{payer16amd,hetzelt2017security}, to modify the memory mapping of a non-colluding service inside a \gls{VM}.
 The modification causes the service to access and return an arbitrary portion of plaintext memory when serving requests.
 This allows an attacker in the \gls{HV} to extract all the \gls{VM}'s main memory in plaintext by repeatedly requesting the same resource and changing its mapping.
 However, the encryption prevents the attacker from locating the \gls{VM}'s most valuable resources in memory prior to extraction, such as keys for \gls{TLS}, \gls{SSH} or \gls{FDE}.
 In the worst case, extracting those secrets requires a full dump of the \gls{VM}'s memory.
 This can take a significant amount of time, depending on the size of the attacker-controlled resource and throughput of the service.
 For example, an extraction speed of about 80 KB/s was reached with web servers providing a resource covering exactly one memory page.
 In this scenario, it takes more than 7 hours and requires 524,288 requests to extract all memory contents of a \gls{VM} with 2 GB of main memory.
 During this time, other clients requesting the same resource also receive arbitrary contents, making full memory extraction conspicuous.

 In this paper, we show that it is possible to overcome these downsides and present an approach that makes \glspl{HV} capable of quickly locating and extracting specific secrets from \gls{SEV}-enabled \glspl{VM}.
 Our approach has two phases, the \textit{observation} and the \textit{retrieval} phase.
 In the observation phase, we exploit the fact that the \gls{HV} is able to observe certain events triggered by \glspl{VM}.
 These \textit{observable events} can, for instance, be page faults which the \gls{HV} handles but also I/O events like network traffic or disk writes.
 We observe and combine such events to identify a minimal set of \gls{VM} memory pages likely to contain the targeted secrets.
 Second, in the retrieval phase, we iteratively extract and analyze the identified set of pages on the fly until we find the targeted secret.
 For this phase, we use the SEVered attack, but could potentially leverage other vectors allowing memory extraction from \gls{SEV}-encrypted \glspl{VM}.

 Our targeted extraction approach offers an inconspicuous, reliable and efficient method to steal various secrets from encrypted \glspl{VM}.
 We demonstrate the potential of our approach by extracting \gls{TLS} and \gls{SSH} keys from a \gls{VM}'s user space memory, and \gls{FDE} keys stored in the \gls{VM}'s kernel space.
 We conduct our experiments on an \gls{SEV}-enabled EPYC processor, running Apache and nginx web servers as well as the OpenSSH server.
 To show that our approach can cope with real-world scenarios where \glspl{VM} can be under varying levels of load, we base our experiments on a load model in which multiple independent clients concurrently access the \gls{VM}'s services during our attack.

\section{AMD SEV and the SEVered Attack}
\label{sec:background}
This section provides background information on AMD \gls{SEV}, the \gls{SLAT} concepts of \glspl{HV}, and the SEVered attack.

\paragraph{SEV}
The AMD \gls{SEV} technology allows for the transparent encryption of main memory of individual
\glspl{VM}.
\gls{SEV} primarily targets server systems and builds on the AMD
\gls{SME} technology, which provides transparent full main memory encryption.
While the goal of \gls{SME} is to protect systems against physical attacks
on the main memory, \gls{SEV} tries to additionally protect memory of individual \glspl{VM} against
attacks from other \glspl{VM} and from a malicious \gls{HV}.
The \gls{SEV} encryption is executed by a hardware AES engine located
in the memory controller.
The keys for the encryption are created and managed by an additional component, the AMD
\gls{SP}.
All keys are ephemeral and never exposed to software on the main CPU.
In contrast to \gls{SME}, \gls{SEV} uses different keys for each
\gls{VM} and for the \gls{HV}.
Additionally, a \gls{VM} running on an
\gls{SEV}-protected system can request encryption and receive proof that
its memory contents are being encrypted, which establishes trust between its owner
and the remote operator.
While \gls{SME} was first integrated into AMD's Ryzen CPUs,
\gls{SEV} was introduced onto the market with the EPYC processor family.
The mainline Linux kernel provides necessary software-level support for \gls{SEV}.

\paragraph{SLAT}
AMD \gls{SEV} integrates with the existing AMD hardware virtualization
technologies marketed as AMD-V. An integral component of hardware
virtualization is an additional address translation, often named
\emph{nested paging} or \gls{SLAT}~\cite{AMD2008nested}. While
non-virtualized systems simply translate virtual addresses directly to physical
addresses, a hardware-virtualized system distinguishes between three different
types of addresses.
When the \gls{VM}
accesses a \gls{GVA}, the guest-controlled first level translation translates
the address to a \gls{GPA}. The \gls{GPA} is then translated to a \gls{HPA} using the
second-level translation controlled by the \gls{HV}.
\gls{SLAT} is completely
transparent to the \gls{VM}. This allows running multiple \glspl{VM} that use the
same \gls{GPA}
space while separating them in physical memory.
With \gls{SEV} enabled, the first level translation from \gls{GVA} to
\gls{GPA} in the encrypted \gls{VM} is non-accessible to the \gls{HV}. But the \gls{HV} is
still responsible for managing physical memory for its \glspl{VM} and is
therefore able to restrict access and change second-level mappings from \glspl{GPA}
to \glspl{HPA}. Since there is no integrity protection in \gls{SEV}, the \gls{HV} can use
\gls{SLAT} to transparently switch a \gls{GPA} to \gls{HPA} mapping to a different
\gls{HPA} page belonging to the same \gls{VM}.

\paragraph{SEVered}

The SEVered attack~\cite{Morbitzer:2018:SSA:3193111.3193112} enables a malicious or compromised \gls{HV} to extract the full memory of \gls{SEV}-encrypted \glspl{VM} in plaintext
by exploiting \gls{SEV}'s missing integrity protection.
SEVered requires a (non-collu\-ding) service in the targeted \gls{VM}, e.g., a web server, offering a remotely accessible resource.
The first step of SEVered is to identify the \glspl{HPA}, i.e., the physical pages, at which the accessible resource is located in the \gls{VM}'s encrypted memory.
The number of pages containing (parts of) the resource depends on the size of the resource as well as on the page size. 
The knowledge about the resource's location allows SEVered to modify the \gls{VM}'s \gls{GPA} to \gls{HPA} mappings to point to arbitrary other \glspl{HPA} of the \gls{VM} instead of to the service's resource.
The modified mapping causes the service to access different memory pages instead of the real resource when handling requests.
In the second step, SEVered repeatedly requests the resource while remapping the memory (using the \gls{SLAT} feature).
This leads to the iterative extraction of an encrypted \gls{VM}'s full memory contents.
The throughput of SEVered depends on the service and on the amount of pages that can be extracted at once.
Our attack uses SEVered for the extraction of main memory from \gls{SEV}-encrypted \glspl{VM}.
Like SEVered, our attack neither requires breaking \gls{SEV}'s cryptographic primitives, nor control over the \gls{SP}.
Likewise, our attack requires control over the \gls{HV}, i.e., a malicious administrator or a compromise of the \gls{HV}.
We refer to~\cite{Morbitzer:2018:SSA:3193111.3193112} for further information about SEVered.

\section{Finding and Extracting Secrets}
\label{sec:extraction}

Our concept for the targeted extraction of secrets from \gls{SEV}-encrypt\-ed \glspl{VM} has two phases:
In the first phase, we start our attack by \emph{observing} the page accesses of the targeted \gls{VM} in the \gls{HV} until an event occurs which indicates the \gls{VM}'s recent use of the targeted secret.
In the second phase, we \emph{search} the \gls{VM}'s memory for the secret by systematically extracting and analyzing the set of observed pages accesses.
This section describes both phases in detail.

\subsection{Observation Phase}

The goal of the observation phase is to narrow down the set of \gls{VM} memory pages possibly containing the targeted secret.
We start the phase at an arbitrary point in time by tracking the \gls{VM}'s page accesses in the \gls{HV} until observing the end of a particular \textit{activity}.
This activity \textit{must} make use of the targeted secret \textit{at least once}.
The start of the activity, denoted by \textit{Activity\textsubscript{Start}}, does not need to be observable by the \gls{HV}.
In contrast, the end of an activity, called \textit{Activity\textsubscript{End}}, \textit{must} be a \gls{HV}-observable event.
This event indicates that the \gls{VM} \textit{recently} used the secret one or multiple times, denoted by \textit{Use\textsubscript{1}..Use\textsubscript{n}}.
As soon as we observe \textit{Activity\textsubscript{End}}, we stop tracking, denoted by \textit{Tracking\textsubscript{End}}.
We do not actively attempt to trigger \textit{Activity\textsubscript{Start}} in order to interfere as little as possible.

To start page access tracking, denoted by \textit{Tracking\textsubscript{Start}}, the \gls{HV} invalidates all the target \gls{VM}'s \gls{GPA} to \gls{HPA} mappings.
As a consequence, each of the \gls{VM}'s page accesses causes an observable event, a \gls{SLAT} page fault.
For each \gls{SLAT} page fault, we record the \gls{GPA} as well as the time and type of the page access (read, write, execute) in a list and re-validate the mapping.
The re-validation \textit{clears} the page from tracking.
This means that each accessed page triggers exactly one page fault and that we track the page \textit{exactly once}, namely the first time it is accessed after \textit{Tracking\textsubscript{Start}}.
The tracking enforces that accesses to the secret will inevitably be recorded.
Note that the secret can be contained in a single page or span over multiple pages and can have multiple occurrences on different pages.

An example for an activity is a \gls{TLS} handshake as part of a request to a web server.
The server uses the targeted secret, in this case its \gls{TLS} private key, to authenticate itself to a client during the handshake.
The \gls{HV} can observe \textit{Activity\textsubscript{End}} by monitoring network traffic, waiting for the packet the \gls{VM} sends to complete the handshake.

\begin{figure}[t]
 \centering
 \includegraphics[width=\linewidth]{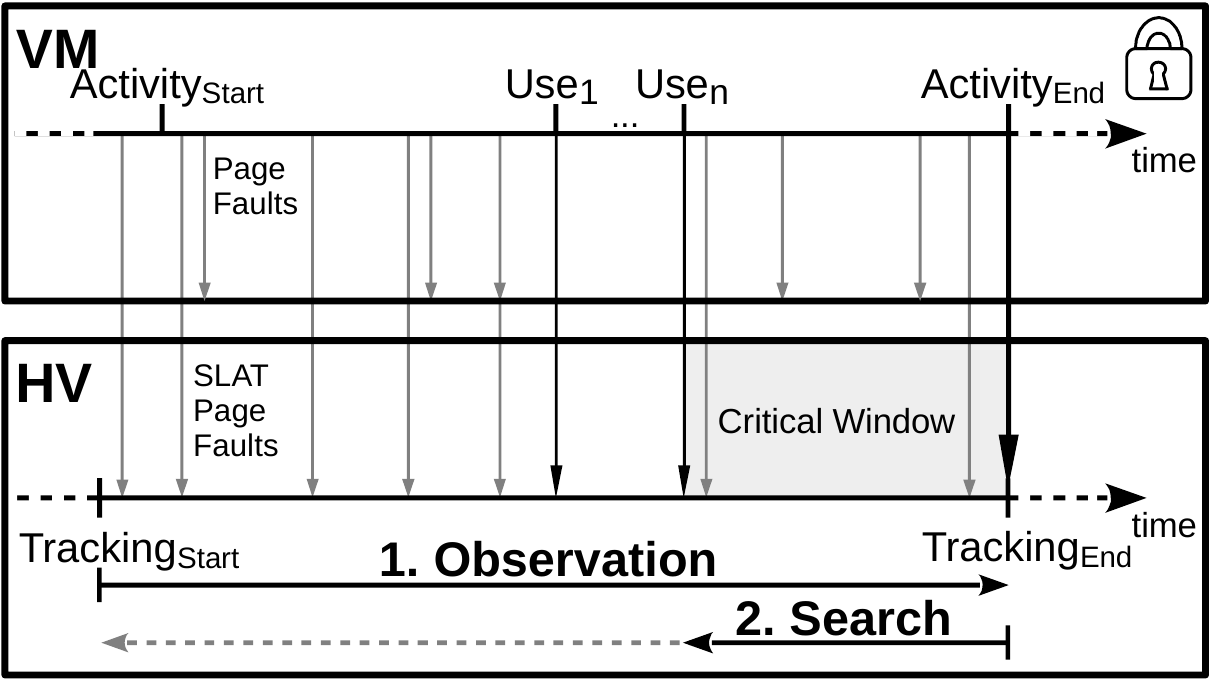}
 \caption{A \gls{HV} first observing an activity inside an encrypted \gls{VM} and then searching for the targeted secret.
 The vertical lines crossing the \gls{VM} boundary into the \gls{HV} box depict the events observable outside the \gls{VM}.}
 \Description[The figure shows a hypervisor observing the behavior of a virtual machine]{
 The illustration depicts an attack scenario with the target virtual machine and the hypervisor in the upper and lower box, respectively.
 The illustration shows the start and end of a virtual machine 's activity along with events triggered by the activity, such as uses of a key.
 Vertical arrows crossing the upper and lower box represent the events observable from the hypervisor.
 These are, for instance, second-level page faults, network packets or disk writes and reads.
 Some of the vertical arrows do not cross the boundary of the virtual machine.
 These are events not observable by the hypervisor, for example, page faults handled by the virtual machine or possibly the start of an activity.
 The illustration clarifies that the search phase starts right after the termination of the observation phase,
 which is when the end of an activity is observed.
 }
 \label{fig:overview}
\end{figure}

\autoref{fig:overview} depicts an attack scenario with the target \gls{VM} and the \gls{HV} in the upper and lower box, respectively.
The illustration shows the start and end of a \gls{VM}'s activity along with events triggered by the activity, such as \textit{Use\textsubscript{1}..Use\textsubscript{n}}.
The vertical arrows crossing the upper and lower box represent the events observable from the \gls{HV}.
These are, for instance, \gls{SLAT} page faults, network packets or disk I/O.
Some of the vertical arrows do not cross the boundary of the \gls{VM}.
These are events not observable by the \gls{HV}, for example, page faults handled by the \gls{VM} or possibly \textit{Activity\textsubscript{Start}}.
Some of the events may be related to concurrent activities, and multiple other activities may potentially make use of the secret as well,
cases which are not depicted in \autoref{fig:overview}.
The illustration emphasizes that \textit{Tracking\textsubscript{End}} concludes the observation phase right after \textit{Activity\textsubscript{End}}.

When starting tracking between \textit{Use\textsubscript{n}} and \textit{Activity\textsubscript{End}},
we do not observe any of the page accesses to the secret.
This means that we are unable to find the secret in the later search phase, requiring to repeat the attack.
This is why we call the timespan between \textit{Use\textsubscript{n}} and \textit{Activity\textsubscript{End}} the \textit{critical window}.
The critical window size is an important factor regarding the quality of the attack.
The smaller the critical window the higher the probability that the attack succeeds.
Further, a small critical window means quick termination of page access tracking after \textit{Use\textsubscript{n}}.
This causes \textit{Use\textsubscript{n}} to be tracked \textit{at the very end} of the phase, and likely only a few more pages to be tracked after \textit{Use\textsubscript{n}}.
We evaluate the critical window size for different scenarios with various levels of load in \autoref{sec:evaluation}.

It is \textit{not} necessary to synchronize the start of the observation phase with a possibly non-observable \textit{Activity\textsubscript{Start}}.
If \textit{Tracking\textsubscript{Start}} takes place \textit{long before} \textit{Activity\textsubscript{Start}}, the observation phase might take longer,
but since every page is tracked only once, this does not lead to a persistent performance impact.
On the other hand, if \textit{Tracking\textsubscript{Start}} takes place \textit{after} \textit{Activity\textsubscript{Start}} (but not inside the critical window),
the tracking period will be shorter and likely output less tracked page accesses.

To conclude, the result of the observation phase is a list of pages in which the page with the targeted secret is contained at least once as long as \textit{Tracking\textsubscript{Start}} is not inside the critical window.
The set of pages in the list is significantly smaller than the whole set of the \gls{VM}'s pages.

\subsection{Search Phase}

The goal of the search phase is to extract the targeted secret from the \gls{VM}'s memory as quickly as possible, i.e., with a minimal number of memory requests.
The input to the search phase is the list of tracked pages acquired during the observation phase.
It is unknown which of the page accesses in the list correspond to \textit{Use\textsubscript{1}..Use\textsubscript{n}}.
The naive extraction of all pages in the list would still require a fairly high number of memory requests to find the secret.
In the following, we describe our approach for a more efficient extraction.

The search phase starts right after \textit{Tracking\textsubscript{End}}, as depicted at the bottom of \autoref{fig:overview}.
We know that \textit{Activity\textsubscript{End}} indicates \textit{recent} use of the secret.
This means that \textit{Use\textsubscript{n}} must have occurred shortly before \textit{Tracking\textsubscript{End}}.
For this reason, we consecutively extract the tracked pages \textit{in backward order} until we find the secret.
We thus start the extraction with the \textit{most recently} tracked pages.
This backward search is shown by the arrow directed to the left at the bottom of \autoref{fig:overview}.
We analyze extracted memory chunks for the presence of the secret \textit{on the fly} to be able to terminate the extraction procedure as early as possible.
On the fly means we search the latest extracted memory chunk for the secret while we request the next chunk.
When finding the secret in the chunk, we terminate the search phase, otherwise we request another chunk.
The actual analysis is specific to the targeted secret and described in \autoref{sec:scenarios} for different secrets.

We propose an optional \textit{preprocessing} step before the extraction to further minimize the number of memory requests.
Preprocessing \textit{filters} page accesses from the list, which \textit{cannot} represent a use of the secret,
and \textit{prioritizes} accesses that are \textit{likely} to represent a use.
The ability to filter and prioritize depends on the use case, in particular, on the specific activity and secret.
In most cases, the secret is a data structure on a page in non-executable memory, allowing to filter all execute-accesses from our list.
The page is likely to be read, but may also be written during an activity.
Depending on the use case, it is also possible that the secret resides in a read-only area, or represents confidential code.
The information about this can often be acquired prior to the attack.
A further possibility for preprocessing is to conduct a representative offline access pattern analysis for the activity to observe the expected timing of \textit{Use\textsubscript{1}..Use\textsubscript{n}}.
An offline analysis is more representative the more the hardware platform and the software configuration inside the \gls{VM} resemble the attack target.
With the gained timing information, an attacker can further filter or re-prioritize pages in the list.

Extracting the secret from the encrypted \gls{VM} using SEVered requires the secret to remain at the same location during the attack.
This means that the secret must not be erased or moved to a different \gls{HPA} by the \gls{VM}'s kernel before the search phase terminates.
We show that the secrets we chose for extraction always fulfilled this requirement and investigate preprocessing possibilities as part of our evaluation in \autoref{sec:evaluation}.

\section{Key Extraction Scenarios}
\label{sec:scenarios}

In the following, we describe the application of our concept for the extraction of targeted secrets at the example of private keys and symmetric \gls{FDE} keys.
We focus on the aspects from \autoref{sec:extraction} that are specific for the type of secret.
These aspects are the activities with their events and the on the fly analysis.

\subsection{Private Keys}
\label{sec:privkeys}

For the extraction of private keys, we focus on the example of web server \gls{TLS} keys.
These keys are resources located in a \gls{VM}'s user space and highly sensitive.
Web servers use these keys to establish authenticated \gls{TLS} channels with clients.
An attacker can make use of a stolen private key for identity spoofing and deceive clients for fraud or data exfiltration.

\begin{description}
\item [Events.] \textit{Activity\textsubscript{Start}} is the start of a \gls{TLS} handshake.
The handshake can be part of an HTTPS request or be directly triggered by a client.
\textit{Use\textsubscript{i}} represents a server's use of the \gls{TLS} key for its authentication during the handshake.
The exact moment of use depends on the key exchange method.
For instance, in case of an \gls{ECDHE}-based key exchange algorithm, this is the moment of signing curve parameters.
For an RSA-based key exchange, this moment is the decryption of the premaster secret encrypted by the client with the server's public key.
\textit{Activity\textsubscript{End}} happens when the \gls{VM} sends the client a specific network packet during the handshake.
We observe these packets with network monitoring tools.
The \emph{change cipher spec} packet is an indicator independent of the specific key exchange algorithm.
Depending on the algorithm, packets sent earlier may be usable indicators as well.
Note that we can also observe or even trigger \textit{Activity\textsubscript{Start}} ourselves in this scenario.
We discuss this aspect in \autoref{sec:discussion}.

\item [On the fly analysis.] The public key and its length are part of the server's certificate and known in advance.
When using RSA, the private components of the key are the factors \textit{p} or \textit{q} of known length dividing the modulus of the public key.
For every extraction request we make, we traverse the extracted chunk of memory and check if it contains a contiguous bit sequence that divides the modulus without remainder.
If so, we found either \textit{p} or \textit{q} and can instantly determine the other factor.
Otherwise, we request the next chunk of memory.
Analyzing a chunk this way usually takes less time than memory extraction with SEVered, see \autoref{sec:evaluation}.

\end{description}
The same approach can be used for extracting \gls{SSH} private keys.
In the \gls{SSH} scenario, the \gls{SSH} server must also use its private key for authentication during the \gls{SSH} handshake when establishing a session.
We evaluate the extraction of \gls{TLS} and \gls{SSH} keys using the Apache, nginx and OpenSSH servers in \autoref{sec:evaluation}.

\subsection{FDE Keys}

The normal approach when using \gls{SEV} is to first perform an attestation of the platform.
The attestation proves to the tenant that the \gls{VM} has been started with \gls{SEV} enabled.
After a successful attestation, the tenant provides the \gls{FDE} key in encrypted form to the \gls{VM}~\cite{AMD2017API}.
This protects the key from eavesdropping adversaries in the network and from being read by the \gls{HV}.
Thereafter, the \gls{FDE} key is present in the \gls{VM}'s memory and can be extracted with our approach.
The \gls{FDE} key is particularly important, because it allows attackers to decrypt the \gls{VM}'s persistent storage gaining access to further valuable secrets.

\begin{description}
\item[Events.] The corresponding activity is a disk I/O operation.
The trigger for \textit{Activity\textsubscript{Start}} is not observable by the \gls{HV} and unlike in the \gls{TLS} key scenario, \textit{Activity\textsubscript{Start}} can have many different triggers.
The trigger can, for instance, be data uploaded to a service, a request to a web server being logged, or an operation of the \gls{VM}'s OS involving disk I/O.
The event \textit{Use\textsubscript{i}} is the \gls{VM}'s use of the \gls{FDE} key to en- or decrypt disk content to be read or written.
We observe \textit{Activity\textsubscript{End}} by monitoring the \gls{VM}'s disk image file in the \gls{HV}.

\item[On the fly analysis.]
We can be sure that we found the secret as soon as we are able to successfully decrypt the \gls{VM}'s persistent storage.
Traversing extracted memory chunks and naively trying each possible sequence as key leads to an inefficient approach.
Our goal is thus to first identify key candidates in extracted memory chunks.
For this purpose, we search the extracted memory for characteristics specific to \gls{FDE} keys based on the following two criteria.

First, the \gls{FDE} key is stored in the \gls{VM}'s kernel in a specific data structure.
This structure has various fields, some of which must have certain value ranges, for instance, kernel addresses pointing to other kernel objects.
Our first criterion for a key candidate is thus the identification of possible \gls{FDE} key structures in extracted memory chunks.

Our second criterion is based on the statistical properties of the \gls{FDE} key.
Because \gls{FDE} is usually AES-based, the kernel derives round keys from the \gls{FDE} key and keeps them in \textit{AES key schedules} in memory.
The round keys have common statistical properties that can be identified with linear complexity.
The first-round key is the AES key itself.
We use aeskeyfind~\cite{keyfind} to search memory chunks for AES key schedules.
Note that candidates that turn out to be false positives are possibly symmetric keys used for other purposes and might also be valuable secrets.
The traversal of memory chunks based on these two criteria takes considerably less time than the extraction of memory with SEVered, see \autoref{sec:evaluation}.
\end{description}
We evaluate the \gls{FDE} key extraction scenario as part of the following section.

\section{Implementation and Evaluation}
\label{sec:evaluation}

In the following, we first define performance indicators and then present our prototype and test setup.
Based on that, we evaluate the extraction of \gls{TLS}, \gls{FDE} and \gls{SSH} keys, as discussed in~\autoref{sec:scenarios}.
In the final part of our evaluation, we present strategies for optimization with preprocessing and summarize our results.

\subsection{Performance Indicators}
The key factors we investigate are the \textit{success probability} and the \textit{attack time}.

\paragraph{Success Probability}
As discussed in \autoref{sec:extraction}, the critical window size is the factor determining the success probability of our attack.
The smaller the critical window, the smaller the probability that the observation phase ends without having tracked the access to the secret.
In our evaluation, we present the success probability for the tested scenarios and provide an upper bound on the size of the critical window.
We call the upper bound the \textit{reaction time} of our attack.
The reaction time is the sum of the critical window (the time frame between \textit{Use\textsubscript{n}} and \textit{Activity\textsubscript{End}})
and the time our prototype requires to detect \textit{Activity\textsubscript{End}} and stop tracking (the time frame between \textit{Activity\textsubscript{End}} and \textit{Tracking\textsubscript{End}}).
\paragraph{Attack Time}
We divide the total attack time of a full attack into its three components: the time required to setup SEVered prior to extraction, the duration of the observation phase and the duration of the search phase.
\begin{description}
 \item[Setup of SEVered.]
The time required to setup SEVered is evaluated in~\cite{Morbitzer:2018:SSA:3193111.3193112} and is thus not subject of our evaluation.
Setting up SEVered usually takes less than 20 seconds, depending on the load of the \gls{VM}.
After setting up SEVered once, we can arbitrarily extract the victim \gls{VM}'s memory and repeat our attack when necessary.
 \item[Observation phase.]
The main factor for the duration of the observation phase is the frequency of the targeted activity.
For instance, a web server under high load will often make \gls{TLS} handshakes while \gls{SSH} logins generally occur less frequently.
 \item[Search phase.]
The duration of the search phase is mainly determined by the amount of memory that has to be extracted until the secret is found.
This is driven by the number of pages we track within the reaction time frame.
The reaction time not only provides an upper bound on the critical window, but also serves as indicator for the expected number of tracked pages.
\end{description}
In our evaluation, we investigate
the number of pages that have to be extracted, and the duration of the observation and search phase.
We call the combined duration of both phases the \textit{attack time}.

\subsection{Prototype and Test Setup}
We implemented our prototype including the functionality required for SEVered based on \gls{KVM}.
To start and stop page tracking and change mappings, we extended the \gls{KVM} API with additional calls, in particular, with \textit{\gls{KVM} system ioctls}~\cite{kvm_api}.
This allows us to launch the attack from user space by communicating with the \gls{KVM} kernel module.
For page access tracking in \gls{KVM} we used the technique from~\cite{Morbitzer:2018:SSA:3193111.3193112,guangrong2016patch}.
While tracking is active, we record all tracked pages in a list in kernel memory.
Upon the call to stop tracking, \gls{KVM} returns the list of tracked pages to user space.

We ran \gls{KVM} on Debian with a page size of 4 KB using an \gls{SEV}-enabled Linux kernel in version \texttt{4.18.13} and QEMU \texttt{3.0.50}.
We used an AMD EPYC 7251 processor with full support for \gls{SEV}.
We created a victim \gls{VM} with 2 GB of memory and one of the four available CPU cores.
We deployed Apache \texttt{2.4.25-3} and nginx \texttt{1.10.3-1} for the \gls{TLS} key scenarios, and Open\-SSH \texttt{7.4} for the \gls{SSH} scenario in the \gls{VM}.
The \gls{FDE} scenario is independent of a service, because the \gls{FDE} key is a kernel resource exclusively used by the OS.
We deployed eleven different web resources on each web server.
We used 4096-bit private keys for \gls{TLS} and \gls{SSH} and a 256-bit symmetric \gls{FDE} key for storage encryption with AES-XTS.
As target for memory extraction with SEVered, we used a page-sized web resource served by nginx.

\begin{figure*}[t]
 \centering
 \includegraphics[width=\linewidth]{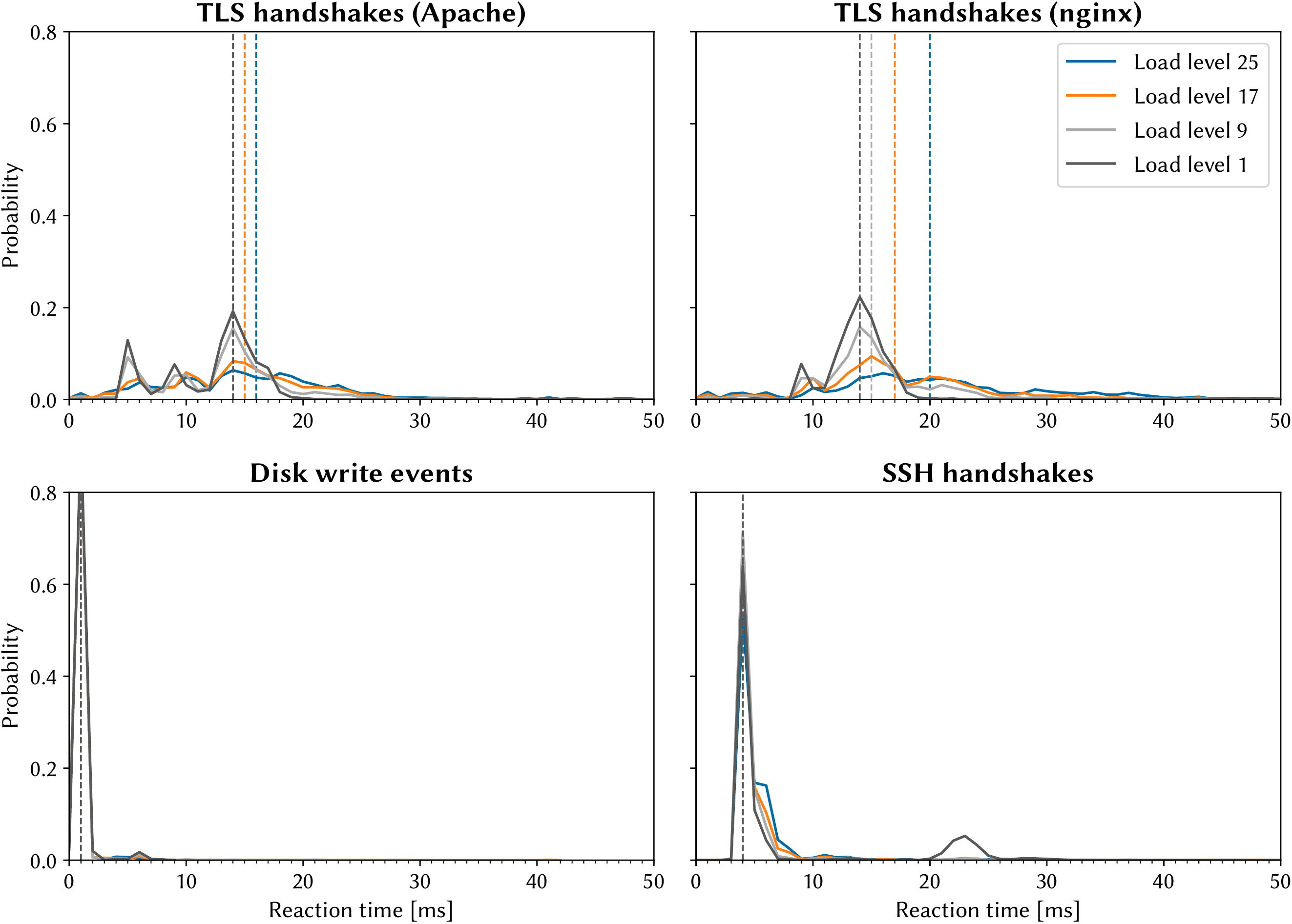}
 \caption{Distribution of the reaction times for all scenarios and load levels.
  The X-axes show discretized time steps of one millisecond and the Y-axes are normalized to one.}
  \Description[The illustration shows four plots with graphs for each evaluated scenario and noise level]{
  The illustration shows the distribution of measured reaction times for each scenario.
  The four graphs in each diagram represent the four load levels.
  The X-axes are discretized in steps of one millisecond, and the Y-axes are normalized to one.
  Vertical dashed lines show the median reaction times over all repetitions for each level,
  providing an upper bound on the median critical window size.
  }
\label{fig:key-access-timing}
\end{figure*}

To capture the handshake messages for the \gls{TLS} and \gls{SSH} key scenarios, we used \texttt{tcpdump} with \texttt{libpcap}, a library for network packet capturing.
For \gls{TLS}, we captured the \textit{change cipher spec} packet the services send to conclude a \gls{TLS} handshake (filter \texttt{tcp[37] == 0x04}).
For \gls{SSH}, we captured the \textit{new keys} message, which concludes the \gls{SSH} handshake (filter \texttt{tcp[37] == 0x15}).
We patched \texttt{libpcap} to execute a system call for \textit{Tracking\textsubscript{Start}} the moment packet capturing begins, and a call for \textit{Tracking\textsubscript{End}} the moment the filtered packet is captured.
This tight interconnection minimizes the reaction time.
To monitor disk I/O events of the \gls{VM} in the \gls{FDE} key scenario, we used the tool \texttt{inotifywait} to observe \textit{inotify} events.
In particular, the \textit{notify} option allows to detect disk writes on the \gls{VM}'s disk image file.
We modified \texttt{inotifywait} to issue the calls for \textit{Tracking\textsubscript{Start}} right before starting to watch events and for \textit{Tracking\textsubscript{End}} as soon as an inotify event is identified.

In real-world scenarios, a tenant's \gls{VM} can show higher or less activity depending on the load caused by its clients.
To simulate this behavior, we executed all our tests based on a \textit{load model} with various \textit{load levels},
representing low to high load.
In our model, a load level of nine, for instance, refers to nine requests per second to the \gls{VM}.
We randomly alternate between the services for each request.
With a probability of $\frac{300}{301}$, we make a request to one of the resources offered by one of the two web servers with equal probability.
With a probability of $\frac{1}{301}$, we initiate an \gls{SSH} login with a user remaining logged in for two minutes.
Compared to the number of web server requests, we execute only few \gls{SSH} logins, as these usually happen less frequently than requests to a web server.
The average duration of the observation phase thus lies in the range of a few seconds to a few hundreds of milliseconds for the web servers
and in the range of a few minutes to tens of seconds for \gls{SSH}.
Note that \texttt{sshd} forks a new process for each new \gls{SSH} connection.
When the session terminates, the process exits and purges its \gls{SSH} key.
This means that the search time must be less than two minutes to extract the \gls{SSH} key before the forked process exits.

We conducted 2,000 independent iterations of our attack for each of the four scenarios on four load levels: level \textit{1}, \textit{9}, \textit{17}, and \textit{25}.
We started our attacks at random points in time while the \gls{VM} processed requests according to the specific load level of our model.
As an initial preprocessing step before the search phase, we filtered all execute-accesses.
In our scenarios, all secrets are data structures located on non-executable memory pages.

\subsection{Success Probability and Reaction Time}

In this part, we investigate the success probability and reaction times.
The four diagrams in \autoref{fig:key-access-timing} illustrate the distribution of measured reaction times for each scenario.
The four graphs in each diagram represent the four load levels.
The X-axes are discretized in steps of one millisecond, and the Y-axes are normalized to one.
The vertical dashed lines show the median reaction times over all repetitions for each level,
providing an upper bound on the median critical window size.

The results for Apache and nginx \gls{TLS} handshakes are depicted in the top row of~\autoref{fig:key-access-timing}.
Both diagrams show a clear peak for the two lower load levels,
indicating a reliable reaction time when the \gls{VM} is not under high load.
For the lowest load level, we can even observe that the reaction time never exceeded 21 milliseconds with Apache and 22 milliseconds with nginx.
For higher load levels, more concurrent activities are executed by the \gls{VM}, and the measurements are more dispersed over time.
Consequently, it becomes likely that more pages have to be extracted in the search phase until the secret is found.
This led to a maximum reaction time of around 50 milliseconds for both nginx and Apache in rare cases.
However, the median reaction time increased to about only 20 milliseconds for nginx, and to about 16 milliseconds for Apache.
As the reaction time is an upper bound for the critical window, the latter is smaller than tens of milliseconds for both Apache and nginx.
We achieved a very high success rate of around 99.99\% for both web servers on all load levels,
meaning that we started tracking inside the critical window only in a few cases.
The high success rate indicates that the upper bound we measured is a very conservative estimate.
This comes from the fact that our prototype requires some time to actually stop tracking and (especially for the \gls{TLS} scenarios) to recognize \textit{Activity\textsubscript{End}}.
Note that if \textit{Tracking\textsubscript{Start}} occurs inside the critical window of a \gls{TLS} handshake,
we still have the chance to observe \textit{Use\textsubscript{i}} of other handshakes being concurrently processed on higher load levels where lots of handshakes are made each second.
The critical window can thus be even smaller on higher load levels.
 
The bottom left diagram in \autoref{fig:key-access-timing} for disk write events shows that our implementation achieved an extremely fast reaction time
of one millisecond in the median for each load level.
Only in a few cases, we encountered a slightly higher reaction time.
In contrast to the \gls{TLS} key scenarios, the behavior was generally independent of the load level.
In the \gls{TLS} key scenarios, the network packets must first be sent by the \gls{VM} to the network interface, on which the \gls{HV} executes more time-consuming network packet capturing.
The interception of disk write events is less complex and introduces less delay.
The success rate for \gls{FDE} key extraction was about 99.99\%, indicating a very small critical window, as confirmed by the upper bound in the graph.

The bottom right diagram in \autoref{fig:key-access-timing} shows that the reaction time for \gls{SSH} handshakes was four milliseconds in the median and mostly independent of the load level.
We encountered only a few samples going up to about 30 milliseconds.
As for the \gls{TLS} scenario, this indicates a small upper bound on the critical window and a possibly quick extraction.
Accordingly, our attack had a success rate of 99.98\%.

\begin{table}[t]
\caption{Statistics for the median length of the observation and search phases,
and for the median number of extracted pages with the median absolute deviation for the different scenarios and load levels.}
\vspace{0.5cm}
\centering
 \begin{tabular}{lccccc}
    \toprule
    \multirow{2}{*}{\pbox{2cm}{\textbf{Use}\\\textbf{Case}}} & \multirow{2}{*}{\pbox{2cm}{\textbf{Load}\\\textbf{Level}}} & \multirow{2}{*}{\pbox{2cm}{\textbf{Median}\\\textbf{Page No}}} & \multirow{2}{*}{\pbox{2cm}{\textbf{~~~~MAD}\\\textbf{Page No}}} & \multicolumn{2}{c}{\textbf{Median Time}} \\
    \cmidrule{5-6} & & & & \textbf{Observ.} & \textbf{Search}\\
     \midrule
     \multirow{4}{*}{\pbox{2cm}{\textbf{TLS}\\\textbf{(nginx)}}}
     & 1 & 102 & 5 & 1.46s & 17.72s \\
     & 9 & 116 & 19 & 0.37s & 15.48s \\
     & 17 & 165 & 69 & 0.32s & 18.61s \\
     & 25 & 301 & 160 & 0.31s & 32.71s \\

     \midrule
     \multirow{4}{*}{\pbox{2cm}{\textbf{TLS}\\\textbf{(Apache)}}}
     & 1 & 128 & 21 & 1.42s & 21.90s \\
     & 9 & 137 & 40 & 0.37s & 17.95s \\
     & 17 & 154 & 80 & 0.33s & 17.44s \\
     & 25 & 171 & 109 & 0.32s & 18.65s \\

     \midrule
     \multirow{4}{*}{\textbf{FDE}}
     & 1 & 70 & 8 & 2.43s & 12.24s \\
     & 9 & 71 & 9 & 2.15s & 9.34s \\
     & 17 & 70 & 8 & 2.08s & 7.84s \\
     & 25 & 69 & 9 & 2.04s & 7.37s \\

     \midrule
     \multirow{4}{*}{\textbf{SSH}}
     & 1 & 7 & 1 & 193.36s & 1.33s \\
     & 9 & 7 & 1 & 27.23s & 0.97s \\
     & 17 & 7 & 1 & 16.19s & 0.83s \\
     & 25 & 7 & 1 & 14.41s & 0.80s \\
    \bottomrule  
 \end{tabular}
\label{tab:stats_measurements}
\end{table}

\subsection{Attack Time}
This part investigates the attack times for each scenario.
\autoref{tab:stats_measurements} summarizes the relevant statistics for the median number of pages to be extracted and the median attack time for every scenario and load level.
For both Apache and nginx, the median number of pages to be extracted until finding the \gls{TLS} key increased between low and high load levels.
We measured an increase of the median from 102 to 301 pages (i.e., 408 to 1,204 KB of memory) for nginx, and from 128 to 171 pages (i.e., 512 to 684 KB of memory) for Apache.
Additionally, the \gls{MAD} increased from 5 to 160 from low to high load for nginx, respectively from 21 to 109 for Apache.
The median number of extracted pages was particularly small compared to the median number of total tracked pages,
which was for both cases between 1,691 and 2,085 (not listed in \autoref{tab:stats_measurements}).
The median duration of the search phase was between about 15.5 and 32.7 seconds for nginx, and between about 17.5 and 22 seconds for Apache.
We measured an average extraction time of around 123 milliseconds for a single page with our SEVered implementation and setup.
We measured this time to fluctuate quite frequently in the scale of a few tens of milliseconds.
This is why a higher median number of extracted pages did not affect the duration of the search phase in a clearly linear way.
The on the fly analysis for a single memory page took about 50 milliseconds.
This means that the page extraction performance is the limiting factor of our attack.
The higher the load, the less time we required for the observation phase, which ranged from 1.46 to 0.31 seconds in case of nginx, for instance.
This is because the probability of quickly observing \textit{Activity\textsubscript{End}} increases with a high frequency of requests.
To summarize, we measured an attack time between about 16 and 33 seconds in the median for the web services.

For the \gls{FDE} key scenario, the amount of pages that had to be extracted was very small and mostly independent of the load level.
Accordingly, the median number of extracted pages was between 69 and 71 for the different load levels (i.e., 276 to 284 KB of memory)
As in the \gls{TLS} key scenario, this number is small compared to the median number of total tracked pages, which was between 2,526 and 3,433.
The overall duration of the search phase was between about 7.4 and 12.3 seconds.
The on the fly analysis for a single memory page took only about 2 milliseconds on average.
We mostly identified the key as part of the AES key schedule, and only occasionally by the kernel data structure, see \autoref{sec:scenarios}.
We measured a slightly decreasing observation time from 2.43 to 2.04 seconds.
This indicates that the \gls{VM}'s OS is regularly writing pages to disk, in our case mostly regardless of the load level.
In sum, the attack time was between less than 9.5 and 14.7 seconds in the median.

In the \gls{SSH} scenario, we merely had to extract seven pages in the median with a \gls{MAD} of one.
This is a particularly small number, especially compared to the median number of 10,102 to 11,094 total tracked pages, omitted from \autoref{tab:stats_measurements}.
We measured a median duration of the search phase of about 0.80 to 1.33 seconds.
This means that the attack works reliably assuming that the \gls{SSH} connection lasts at least 1.33 seconds.
Similar to the \gls{TLS} key scenario, the on the fly analysis of a memory page took about 50 milliseconds.
With our load model, the observation time of about 14 to 194 seconds was comparably high for the \gls{SSH} scenario.
This is another reason why the number of extracted pages was especially low for the \gls{SSH} case.
In long observation phases, we already tracked a high number of pages before \textit{Use\textsubscript{n}},
making it very unlikely that many pages are tracked within in the reaction time frame at the end of the activity.

\subsection{Optimization with Preprocessing}
As discussed in~\autoref{sec:extraction}, preprocessing with prioritization and filtering is an optional optimization before the search phase.
Preprocessing usually requires a priori knowledge about the use case and behavior of the \gls{VM}, which may not always be available.
This behavior may also vary between different hard- and software configurations.
For our evaluation, we already used the knowledge that all secrets are data structures located on non-executable memory pages.
This allowed us to filter execute-accesses from the list of tracked pages.
The amount of pages to be extracted was thereby reduced by about 22\% on average over all samples.

\textit{Use\textsubscript{n}} was a read-access in 96\% of our attacks for the \gls{TLS} handshakes and in 93\% a write-access for the \gls{SSH} handshakes.
For disk write events, \textit{Use\textsubscript{n}} was always a read-access.
Whether the page containing the secret is tracked as read- or write-access depends on the other data located within the page.
The type of access thus cannot be predetermined with certainty.
Filtering of write-accesses could significantly reduce the attack time, but could also reduce the success probability.
Also, prioritizing the extraction of read-accesses over write-accesses in the list would boost the attack in most cases, but could also introduce costly outliers.

Another possibility for prioritization is knowledge about the reaction time, as shown in \autoref{fig:key-access-timing}.
The graphs for the two web server scenarios show that the reaction time was rarely less than eight milliseconds before \textit{Tracking\textsubscript{End}}.
Re-arranging these early accesses further back in the list of tracked pages can thus reduce the amount of pages to be extracted until the secret is found.
The same observation can be made for the \gls{SSH} scenario, where {Use\textsubscript{n}} never happened less than three milliseconds before \textit{Tracking\textsubscript{End}}.
However, in this case the number of pages to be extracted is already so small that further optimization may not be required.

The reaction times in \autoref{fig:key-access-timing} can also help to determine a good criterion for restarting the attack when a secret has not been found after a certain number of extracted pages.
For instance, page accesses tracked later than 30 milliseconds before \textit{Tracking\textsubscript{End}} are likely exceeding the reaction time frame and thus unlikely to be a candidate for {Use\textsubscript{n}}.
This can be used as a criterion to detect that \textit{Tracking\textsubscript{Start}} was inside the critical window and to restart the attack from the observation phase.

\subsection{Summary}
For all evaluated scenarios, both performance indicators are very promising.
We found that the critical window was very small in all cases.
\textit{Tracking\textsubscript{Start}} was thus inside the critical window only in a few cases, resulting in a very high success probability throughout all scenarios and load levels.
The most important factor for the attack time, the duration of the search phase, was also very small.
Extracting all memory from our \gls{VM} with 2 GB of main memory would take more than 7 hours with SEVered~\cite{Morbitzer:2018:SSA:3193111.3193112}.
Assuming that a key can reside not only on one but on several pages, the naive extraction would require several hours on average to find the key.
Our approach can extract secrets faster by several orders of magnitude.

In cases when \textit{Tracking\textsubscript{Start}} is inside the critical window,
the attack fails and we extract all tracked pages without finding the secret.
In such cases, we have to repeat both the observation and search phase.
To avoid a lengthy extraction of all tracked pages in unsuccessful attempts, the search can be canceled early when the likelihood of finding the secret drops,
according to our evaluated distribution.
For the following search phase, all pages extracted in the previous attempts can then be excluded from extraction given that the secret does not change its location.
In sum, our results have shown that our prototype is able to quickly and reliably extract different sensitive secrets and performs well even under high load.

\section{Discussion}
\label{sec:discussion}

In the following, we discuss further important aspects of our attack and possible countermeasures:

\paragraph{Overhead} The overhead caused by the tracking itself is limited, because each accessed page only triggers a \gls{SLAT} page fault once.
We neither detected perceivable effects like delays in response times in the \gls{HV} nor inside the \gls{VM}.
The host system and \gls{VM} remained stable even on the highest load level.
We measured only a small additional delay of web and \gls{SSH} server responses when tracking was active.

\paragraph{Low Memory} When the \gls{VM} is low on memory, its kernel might try to free memory by swapping out pages, by unmapping file-backed pages,
or by killing processes.
A page containing the secret might then be re-used by another process or by the kernel during our attack.
In such a case, we are still able to extract the memory contents of the page, but its contents might have already been overwritten.
We did not encounter such cases in our tests.

\paragraph{Triggering Activities} In our concept, we start tracking at an undefined point in time and do not actively trigger activities to interfere as little as possible
with the \gls{VM}'s normal operation.
Our concept worked well in our evaluated scenarios, because we extracted frequently used secrets.
In the \gls{SSH} scenario, however, the key may be used rather infrequently.
This is, for instance, the case when an administrator logs in to a web server for maintenance only from time to time.
When a secret is rarely used but the attacker requires the observation phase to be as short as possible,
the attacker can consider the active triggering of an activity.
In the \gls{SSH} scenario, an attacker can actively start a login procedure without a user account.
\gls{SSH} servers use their key for server authentication and wait for the user to authenticate with a default timeout of two minutes.
An attacker can thus initiate a login and extract the \gls{SSH} key before the session timeout without waiting for a legitimate user to login.
Note that active triggering might increase the probability of the attack being detected and might not always be possible.

\paragraph{Portability} We expect that our approach can be transferred to other scenarios and configurations than evaluated.
Our approach does not depend on specific service or library versions.
Furthermore, our approach is not tied to a specific \gls{SEV} processor and mostly independent of the \gls{VM}'s performance and OS.
Our approach can also be leveraged to extract other types of memory, such as confidential code, documents or images.
The performance of our approach can differ on systems with other hard- and software configurations.
However, we expect the performance to vary only slightly assuming that \textit{Tracking\textsubscript{End}} can be observed quickly.
We ran several tests in which we assigned our \gls{VM} more memory, multiple cores and in which we configured the web servers to utilize a high number of worker processes.
The performance indicators remained coherent with our evaluation results in all runs.

\paragraph{Countermeasures}
A countermeasure against our attack is to prevent the SEVered attack~\cite{Morbitzer:2018:SSA:3193111.3193112}, which we rely on for memory extraction.
Further, our attack relies on targeted secrets to remain at their memory location during our attack.
Purging secrets in memory after use would cause the search phase to fail.
We found \gls{TLS} and \gls{FDE} keys to always remain at their memory location in our tests.
However, in case of \gls{SSH} keys, the processes forked by the \gls{SSH} daemon for initiating new \gls{SSH} connections purge their private key when a session terminates and then exit.
This means that an \gls{SSH} session must remain open until the secret is extracted.
This, for instance, requires a user to remain logged in or a login attempt to remain pending over the time of the search phase, which is less than 1.5 seconds in our case.

Systematically purging all sorts of secrets from main memory after use would require adapting existing software.
For some secrets, purging might not be feasible.
An example is the \gls{FDE} key, which is constantly required for disk I/O.
A more promising solution is to relocate the most valuable secrets from main memory to dedicated hardware.
Since SEVered can only extract contents from main memory, storing secrets in hardware would prevent them from being extracted by a malicious \gls{HV}.
This can, for example, be realized using \glspl{HSM}.
Additionally, hardware-based disk encryption can be used to protect the \gls{FDE} key.

\section{Related Work}
\label{sec:related_work}
While there are established \gls{VMI} frameworks \cite{libVMI,volatility,rekall} for data analysis and extraction on unencrypted \glspl{VM},
the systematic extraction of memory from encrypted \glspl{VM} has not been subject to extensive study.
On AMD \gls{SEV} platforms, the \gls{SP} protects page encryption and the corresponding keys from the \gls{HV}.
This makes it infeasible to directly read memory contents from \gls{SEV}-enabled \glspl{VM} as long as the \gls{SP} cannot be compromised \cite{safefirmware}.
Payer~\cite{payer16amd} early discussed the missing integrity protection on AMD \gls{SEV} platforms.
By remapping memory in the \gls{HV}, this can be used to extract memory without compromising the \gls{SP}, as done by the SEVered attack \cite{Morbitzer:2018:SSA:3193111.3193112}.
While SEVered allows the extraction of data, it does not provide concepts for quickly extracting specific secrets.

Buhren et al. presented an attack \cite{hetzelt2017security} to gain remote code execution with user privileges on an \gls{SEV}-enabled \gls{VM}.
Their approach exploits memory remapping to modify the control flow of an \gls{SSH} service.
The first step is an off-line tracing of the system call sequences performed during an \gls{SSH} login on a comparable, unencrypted \gls{VM}.
The goal of this analysis is to determine the behavior of a \gls{VM} accessing the login information of the \gls{SSH} session, the \textit{credentials data structure}.
The next step is to wait for a victim user to login to the \gls{SSH} service.
With the information gained in the off-line analysis, they identify the memory page containing the user's login information.
They then try to illicitly login by remapping the valid user's credential data structure to the one the \gls{SSH} service creates during the illicit login attempt.
This allows the attacker to re-use the victim user's login information.
In their evaluation, they achieved a success rate of around 23\%.
The low rate was primarily caused by the fact that the \gls{SSH} service may store the \textit{credentials data structure} at different offsets within the page.
As a condition for a successful attack, the \gls{SSH} service must have stored both the victim user's and attacker's credentials data structures at the same page offset.
Besides being quite invasive, this approach requires access to a comparable \gls{VM}, detailed analysis of the \gls{SSH} service, user interaction,
and data being incidentally stored at specific offsets.

The attack described in \cite{2017arXiv171205090D} follows the same goal of gaining remote code execution on an \gls{SEV}-encrypted \gls{VM}, but does not exploit remapping.
Instead, the authors describe a \textit{ciphertext block move} attack, which also exploits the missing integrity protection.
The authors argue that it is possible to move memory contents in physical memory.
This is because the \gls{HPA} is not part of the AES-based encryption scheme itself but is incorporated into the encryption result in a later step
with a reversible physical address-based tweak algorithm that uses static parameters.
After reversing the tweak, ciphertext can be moved and the tweak re-applied with the target \gls{HPA}.
The authors describe a method that moves the pages to exploit an \gls{SSH} process.
Both the approaches in \cite{hetzelt2017security} and \cite{2017arXiv171205090D} were, to the best of our knowledge, not confirmed on real \gls{SEV} hardware.
The \textit{ciphertext block move} attack could possibly be leveraged for the memory extraction as an alternative to the remapping in SEVered.

On the side of defenses, Fidelius \cite{8327028} is a software-based extension to \gls{SEV}.
This extension is a privileged module separate from the \gls{HV} that restricts the \gls{HV} from accessing specific critical resources with \textit{non-bypassable memory isolation}, for instance, to prevent replay attacks.
The authors provide a \gls{VM} lifecycle concept that describes how to start Fidelius and provide tenants proof that the system runs Fidelius in addition to \gls{SEV}.
This requires trusting the Fidelius module instead of the operating \gls{HV}.

Intel announced the implementation of its own hardware-based memory encryption approach called \gls{MKTME} \cite{intel_mktme}.
According to our understanding of the specification, \gls{MKTME} does not protect from a malicious or compromised \gls{HV}, but only from memory attacks from outside.
The \gls{HV} remains, for instance, capable of enabling or disabling the encryption, or to handle the sharing of memory with other \glspl{VM}.

\section{Conclusion}
\label{sec:concl}

We presented an approach for the efficient extraction of secrets from \gls{SEV}-encrypted \glspl{VM}.
Compared to time-consuming, naive memory extraction, our two-phased approach exfiltrates secrets unobtrusively and quickly with a high success probability.
In the first phase, we track the page accesses of an encrypted \gls{VM} until detecting an event indicating that the \gls{VM} recently accessed the secret.
In the second phase, we leverage an existing attack for memory extraction to systematically retrieve the tracked pages and simultaneously analyze their contents to quickly identify the secret.
We presented various use cases for highly sensitive secrets usually found in \glspl{VM} in cloud scenarios.
We performed an evaluation for these cases on a fully \gls{SEV}-enabled EPYC processor with varying levels of load, usually caused by independent clients not involved in the attack.
Our results show that we are able to extract \gls{TLS} keys after a handshake in less than 15.5 seconds in the median on lower load levels and in no more than about 32.7 seconds in the median on our highest evaluated load level.
The extraction of the \gls{FDE} key after a disk write event took between less than 7.4 seconds and 12.3 seconds in the median.
The extraction phase for \gls{SSH} keys after an \gls{SSH} handshake took about 0.8 to 1.35 seconds in the median.
We expect that our approach can be used for the extraction of further types of secrets, which we are going to investigate in future work.

\section*{Acknowledgments}
This work has been partially funded in the project CAR-BITS.de by the German Federal Ministry for Economic Affairs and Energy under the reference 01MD16004B.
We would like to thank Michael Velten for the implementation of the tool that searches extracted memory dumps for the private components of public key moduli, see \autoref{sec:scenarios}.